\documentclass[twoside]{article}
\usepackage[a4paper]{geometry}
\usepackage[utf8x]{inputenc}
\usepackage[T1]{fontenc}
\usepackage{lmodern}
\usepackage{fRReE}
\usepackage{url}
\usepackage{hyperref}
\usepackage{natbib} 
\usepackage{graphicx}
\usepackage{float} 

\RRlogo{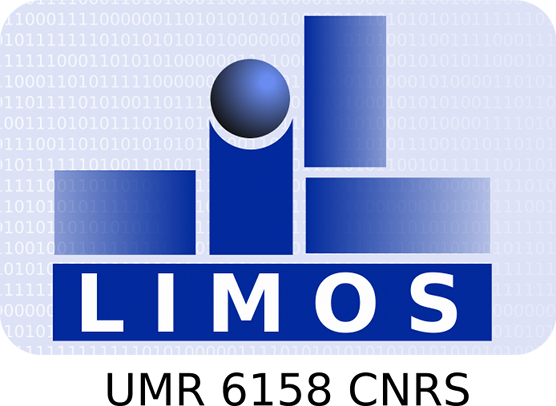}

\RRrc{limos}
\RRrcname{LIMOS - UMR CNRS 6158}
\RRrcaddr{
Campus des C\'ezeaux\\
B\^atiment ISIMA\\
BP 10125 - 63173 Aubi\`ere Cedex\\
France}
\RRrcwebsite{\url{http://limos.isima.fr/}}

\RRdate{October 2010}

\RRauthor{
Jonathan \textsc{Passerat-Palmbach}\RRaffiliation{This author's PhD is partially funded by the Auvergne Regional Council and the FEDER}
\RRaffiliation[sfn0]{ISIMA, Institut Sup\'erieur d'Informatique, de Mod\'elisation et de ses Appplications, BP 10125, F-63173 AUBIERE}
\RRaffiliation[sfn1]{Clermont Universit\'e, Universit\'e Blaise Pascal, LIMOS, BP 10448, F-63000 CLERMONT-FERRAND}
\RRaffiliation[sfn3]{CNRS, UMR 6158, LIMOS, F-63173 AUBIERE}
   \and
\\Claude \textsc{Mazel}
\RRaffiliationref{sfn0}
\RRaffiliationref{sfn1}
\RRaffiliationref{sfn3}
   \and
\\Antoine \textsc{Mahul} 
\RRaffiliationref{sfn0}
\RRaffiliationref{sfn1}
\RRaffiliationref{sfn3}
   \and
\\David R.C. \textsc{Hill}
\RRaffiliationref{sfn0}
\RRaffiliationref{sfn1}
\RRaffiliationref{sfn3}
}

\RRauthorhead{Passerat-Palmbach et. al}
\RRtitle{Reliable Initialization of GPU-enabled Parallel Stochastic Simulations Using Mersenne Twister for Graphics Processors}
\RRsource{European Simulation and Modeling Conference 2010}
\RRtitlehead{ESM 2010}

\RRcopyright{\textcopyright 2010 Eurosis}
\RRoriginalpages{187-195}
\RRadditionalinformation{ISBN: 978-90-77381-57-1}

\RRabstract{
Parallel stochastic simulations tend to exploit more and more computing power and they are now also developed for General Purpose Graphics Process Units (GP-GPUs). Consequently, they need reliable random sources to feed their applications. We propose a survey of the current Pseudo Random Numbers Generators (PRNG) available on GPU. We give a particular focus to the recent Mersenne Twister for Graphics Processors (MTGP) that has just been released. Our work provides empirically checked statuses designed to initialize a particular configuration of this generator, in order to prevent any potential bias introduced by the parallelization of the PRNG.

}
\RRkeyword{Parallel Stochastic Simulation; Random Numbers Generation; GP-GPU; CUDA; MTGP; TestU01}

\begin{document}
\makeRR   

\section{INTRODUCTION}
\paragraph*{}Over the past five years, simulationists have leant towards Graphics Process Units (GPUs) to compute the heavy tasks bound to their research activity. Stochastic simulations, especially Monte Carlo \citep{Gentle2003}, are widely spread through many scientific communities. General-Purpose GPUs (GP-GPUs) with many cores are very interesting for this simulation technique. Depending on the domain in which this kind of simulation is used, results may suffer from a weak parallelization technique with some stochastic streams of poor quality. We have shown that for sensitive research domains, like nuclear medicine, parallel simulations cannot cope with biased results \citep{ElBitar.etal.2006, Reuillon.etal.2008}.

\paragraph*{}Finding a fast and reliable Pseudo Random Number Generator (PRNG) to feed a sequential stochastic simulation is not a problem for many application domains since more than a decade. This issue has been tackled in many reference studies for CPU based PRNG \citep{LEcuyer1990}. \citep{Park.Miller1988} raised the fact that one should consider the couple (application, PRNG) instead of limiting the study to the intrinsic qualities of the PRNG. For instance, a very good generator like Mersenne Twister (MT) \citep{Matsumoto.Nishimura1998} should not be trusted for a cryptographic application and the initialization of such generators has to be done carefully. Indeed, for some years, this very nice generator was sensible to its initialization status. Even if we never have a universal generator, the MT family of generators \citep{Saito.Matsumoto2008}, the WELL generators \citep{Panneton.etal.2006} and some advanced Multiple Recursive pseudo-random numbers Generators (MRGs) from L’Ecuyer \citep{LEcuyer.etal.2002} give very good results when considered for parallel computing in a wide range of applications.

\paragraph*{}If some criteria have been gathered by Coddington \citep{Coddington1996} for sequential and parallel simulation to characterize good PRNGs, it is often safer to refer to empirical testing software. Several libraries have been designed in this way, the oldest and most renowned include the statistical tests proposed by Knuth \citep{Knuth1969}, the DieHard testing suite designed by Marsaglia \citep{Marsaglia1996} and the Brown’s DieHarder suite \citep{Brown.etal.2010}. The TestU01 library provided by L’Ecuyer \citep{LEcuyer.Simard2007} is the ultimate test software at the time of this paper writing. The latter contains a battery of tests called BigCrush, the most stringent set of tests that a PRNG can be faced nowadays.

\paragraph*{}The main problem we are still facing today is to ensure the correct behavior of the PRNG when distributed across a tight or large coupled computing architecture. The literature currently provides quite a few references about stochastic streams distribution on classical hardware architectures \citep{Mascagni1997}, \citep{Traore.Hill2001}, \citep{Bauke.Mertens2007}, \citep{Hill2010}. The set of references is even poorer when considering a GP-GPU environment. Indeed, restricted parallel hardware architectures like the Single Instruction Multiple Data (SIMD) family, which GPUs belong to, do highly impact the implementation of generators. In addition, we still have to select the best way to allocate random substreams to these manycore architectures.

\paragraph*{}In this paper, we bring our contribution to these problems:
\begin{itemize}
\item{We survey the current PRNGs implementations and solutions available to distribute stochastic streams through a GPU architecture.}
\item{The parameterized PRNGs family is proposed taking the example of MTGP \citep{Saito2010}, the generator we find the most interesting in this context.}
\item{We present the latter and discuss its quality with empirical benchmarks and statistical analysis. Reliable initialization data structures for a particular configuration of MTGP are also presented.}
\end{itemize}

\paragraph*{}In this manner, we first intend to analyze the features of a particular generator designed for GPU hardware architectures: MTGP. The second purpose of this study is to give reliable parameters to initialize this PRNG, without introducing any potential bias in the parallel stochastic simulations based upon it.

\section{PSEUDO RANDOM NUMBERS FOR GP-GPUs}
\paragraph*{}Until recently, designing a PRNG for GPU-enabled platforms could be very tricky as it forced programmers to deal with graphics Application Programming Interfaces (APIs). Some implementations are presented in \citep{Sussman.etal.2006}. The authors especially list the limitations of these GPU dedicated PRNGs due to the past weaknesses of the hardware. Limited output per thread or untruthful operations were part of the restrictions that made these PRNGs feeble for High Performance Computing (HPC) applications. Consequently, a common way to deal with random numbers on GPU was to generate them on CPU before transferring them on the graphics processor. This solution has to face the well-known bottleneck of data transfer between the CPU host and the GPU device. Even with nowadays PCI Express 16X running at 8GB/s, this remains a challenge for high performance applications.

\paragraph*{}Since 2008 and the recent advances from NVIDIA, new GPU software and hardware architectures offer the precision and speed needed for many HPC applications. Now, PRNGs can be directly implemented into the GPU. Recent works propose this new kind of generators. Langdon presents a minimal implementation of the standard Park Miller PRNG \citep{Park.Miller1988} on a NVIDIA 8800 GTX GPU in its paper from 2008 \citep{Langdon2008}. He announces a speed up of more than 40 compared to his Intel 2.40 GHz CPU. One year later, \citep{Langdon2009} increased again the speed of his application by four by using the new NVIDIA technology CUDA (Compute Unified Device Architecture) \citep{NVIDIA2010} with a Tesla T10 GPU. Nevertheless, we do not advise the use of this old generator which has many known flows, though it was still in use until recently in some well distributed networking simulation software \citep{Entacher.Hechenleitner2003}.

\paragraph*{}CUDA has been designed to allow developers to easily harness the computation power of GPUs. In his first implementation, Langdon had to deal with a complex and unadapted graphics API. With CUDA, developers can program GPUs without wasting their time making algorithms and their data fit into graphics dedicated data structures, such as pixels shaders. Furthermore, CUDA does not propose a new programming language but only some C extensions, making it easier to learn for C familiars. The CUDA appellation also refers to the name of the new NVIDIA GP-GPU architecture. This generation of graphic boards tries to fulfill the requirements noted in the conclusion of the previously cited \citep{Sussman.etal.2006}, with for instance an implementation of the IEEE 754 floating point numbers standard. The new generation of boards based upon the Fermi architecture is now proposing configurable L1 cache, ECC memory and a considerable increase of performance in double precision, while owning twice as much cores as the antecedently mentioned T10 processor.

\paragraph*{}Although these highly parallel devices bring much more peak performances than CPUs, they must be carefully programmed to deliver the expected power. In fact, GPU architectures combine a manycore approach with SIMD vector cores. As vector processors do, GPU-enabled algorithms need to repeat the same operation on different data to correctly exploit the device. This is the main reason of the recent dedicated PRNGs proposals. In 2006, \citep{Saito.Matsumoto2008} proposed an SIMD version of the famous ‘Mersenne Twister’ called SIMD-oriented Fast Mersenne Twister (SFMT). Although this algorithm can be used on regular CPUs or on SIMD enabled CPUs (using either SSE or AltiVec vector instructions), it cannot be directly transposed to a GPU architecture. Most PRNGs have to be rethought from scratch to leverage GPUs characteristics, once again, we always have to take into account the target application. In the case of a CUDA implementation, these couples are well surveyed in \citep{Howes.Thomas2007}.

\paragraph*{}Given that CUDA defines software levels that map the device architecture, PRNGs implemented using this technology can be organized at one of the following scopes, corresponding to the main elements of the CUDA framework: a thread, a block of threads or a kernel (the program running on the whole GPU). All these approaches have been studied in the literature. In \citep{Zhmurov.etal.2010}, authors present three basic generation algorithms working either with a single instance of the PRNG for the kernel or with an instance per thread. The three algorithms exposed are quite basic: Ran2, Hybrid Taus and a Lagged Fibonacci generator. In the same way, \citep{Langdon2009} chooses to generate a number per thread in its GPU version of the Park-Miller algorithm. The last strategy is proposed in \citep{Saito2010} where a new variant of the MT algorithm spreads independent PRNGs through each thread block, thanks to an algorithm known as Dynamic Creator (DC) \citep{Matsumoto.Nishimura2000} that we will detail later.

\paragraph*{}Beyond the nature of the PRNG algorithm, we prefer to focus on the scope chosen for each implementation. Indeed, we formerly insisted on the need to consider the target application and the PRNG as a pair. Obviously, new PRNG algorithms have to take advantage of GPU intrinsic properties such as heterogeneous memories, or thread organization. The former highly impacts the PRNG performance. Considering the  approach using a generator per thread approach, an internal state array has to be saved in each thread. CUDA related works like \citep{Kirk.Hwu2010} specify that arrays declared for a thread are stored in the local memory, implemented in RAM. Equivalently, with a PRNG for all thread approach, the global memory is solicited to store the state of the PRNG. Each thread draws a number and updates its component of the state in global memory, implemented in RAM too. These two approaches make a heavy use of global memory, which has the advantage to be persistent across kernel launches within the same application. Yet, this RAM area is quite slow, it implies a 400 to 800 clock cycles latency because it is not cached \citep{NVIDIA2010}. So, even if the global memory storage is compulsory to save the PRNG state between two kernel calls, one can use the shared memory, reachable by every thread within a block, to manipulate PRNG data. Indeed, it is implemented on-chip and is consequently as fast as registers. A good example of this choice is the "to be published" paper introducing MTGP \citep{Saito2010}.

\paragraph*{}In our opinion, a good GPU PRNG should employ shared memory. A PRNG per block approach seems to be the most appropriate way to implement a source of randomness, first, because it exploits the quickest memory, second, for the sake of applications upgradability. Since hardware architectures evolve very quickly, we cannot afford to rethink algorithms every time the memory amount or number of threads available doubles. So, fixing a block of threads grained scope for a PRNG algorithm is the safest solution to eschew lots of modifications tied to frequent hardware evolutions. This is the reason why we have decided to study in details the Saito proposition: MTGP.

\section{DESCRIPTION OF MTGP}
	\subsection{Data Structure}
\paragraph*{}At the time we are writing this paper, MTGP was not referenced by any scientific publications yet, except that we found on the Internet that Saito's work is to be published soon \citep{Saito2010}. We must describe its features in order to introduce the goal of our study.
\paragraph*{}First of all, MTGP is obviously inheriting from the properties of its elder, though it is quite different from a simple GPU implementation of the original MT, as seen in \citep{Podlozhnyuk2007}. As a matter of fact, Saito uses the original paper describing MT \citep{Matsumoto.Nishimura1998} to lay his generator out. Thus, we will see that MTGP can suffer from some little problems already spotted for the MT "family". Since we often champion this family of generators, \citep{Reuillon2008} has studied $2^{16}$ statuses of the original MT algorithm using the TestU01 Crush test battery from \citep{LEcuyer.Simard2007}. The involved tests verify the linear complexity of the random sequence. MTGP is based upon the same linear recurrence to create random sequences, so it is not recommended for cryptographic purposes.

\paragraph*{}We have also noted that MTGP specified a common notion of the generators belonging to the parameterized family. We distinguish this cast of generators by the compound form of their data structures. It contains two distinct elements implied in the generation algorithm, we call them seed status and parameterized status. The first is basically the common seed given by the user to initialize a generator. The second stores parameters determined at a particular PRNG creation, it is supposed to be sort of a unique signature of the generator. Both these concepts were already present in MT. MTGP makes them more precise by explicitly using a data structure of the form we described in this paragraph. We propose a simple class diagram of this concept in figure \ref{status_class_parameterized_PRNG}:
 
\begin{figure}[!h]
	\centering
	\includegraphics[keepaspectratio, scale=0.5]{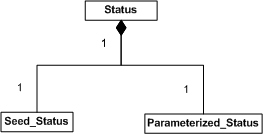}
	\caption{Class diagram of a parameterized PRNG}
	\label{status_class_parameterized_PRNG}
\end{figure}
\vspace{-0.2in}

\paragraph*{}MTGP takes this idea one step further by introducing two kinds of statuses: the references and the fasts. The latters use pre-stored elements to decrease the initialization time, and programming techniques such as inlining to speed up the execution time. However, it results in a memory greedier status.

\paragraph*{}Parameterized statuses are a common way to ensure independence between parallel stochastic streams. This technique is called parameterization in the literature \citep{Hill2010}. Depending on the way it is settled, it can lead to poor results \citep{Matteis.Pagnutti1995}. Unfortunately, we do not have any mathematical theorem allowing us to check the independence between two generators, according to their parameters. However, MT came along with the DC algorithm, designed to create large sets of independent generators. This algorithm integrates an identifier, often the one of the processor or thread that will host the PRNG, to distinguish parallel random streams. It uses it as a part of the characteristic polynomial of the matrix used by the MT algorithm. We can easily conclude that if characteristic polynomials are prime to each other, their associated matrices will also be unique in the same context. Hence, we obtain independent parameter sets.

\paragraph*{}Hopefully this algorithm has been renewed for MTGP, enabling us to proceed in the same way. Furthermore, it improves the original algorithm by allowing the user to get a larger set of $2^{32}$ parameters, where the original algorithm could only deliver $2^{16}$ sets. This number is now too small for large computing grids such as the European Grid Initiative (EGI), with more than 240000 cores at the time we are writing this paper. This latter point forces us to keep skeptical concerning the independence of the MTGP produced by the new DC. Saito explains that a SHA1 (Secure Hash Algorithm) checksum of each characteristic polynomial is generated to let the user check he did not get duplicated entries.

	\subsection{Architecture Independence}
\paragraph*{}One of the most interesting features of MTGP is to be available for both CPUs and GPUs architectures. Even if MTGP has been designed to run on GPUs, you can also find a CPU version at the same location \citep{Saito2010a}. We have based our study on the capability of the generator to merge transparently in CPU-based applications. Hence, we were able to test the PRNG on any CPU-based host with reliable and well-known tools like TestU01. This way, we have avoided the hazardous implementation of a new empirical test battery, which would have to be validated before.
\paragraph*{}Moreover, this property is really precious in our opinion. We intend to use this PRNG for stochastic simulations following the hybrid computing paradigm, where the sequential part of the application runs on a CPU host, while the parallel-one is executed on a GPU board. In such cases, stochastic streams will furnish random numbers to both the CPU and the GPU. With a PRNG like MTGP, we can keep our simulations homogenous, using the same PRNG on each computing element engaged. Handling independent parallel stochastic streams becomes understandably important when you have to deal with such hardware configurations. We will study in a next paragraph the specific DC coming along with MTGP to maximize this independence.
\paragraph*{}Now considering the new elements we introduced in this whole section, we can extend our previous object model to the particular MTGP. Figure \ref{class_components_MTGP} depicts its main components: 

\begin{figure}[H]
	\centering
	\includegraphics[keepaspectratio, scale=1.5]{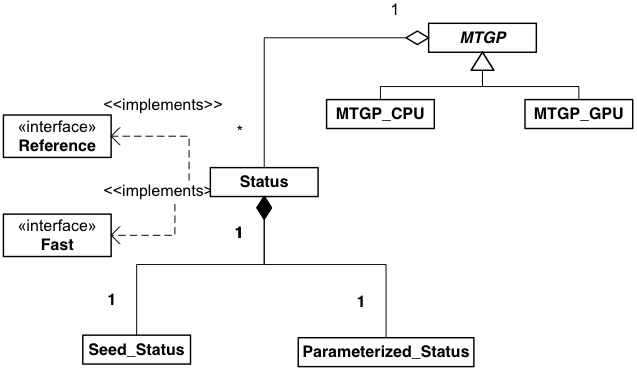}
	\caption{Class Diagram for MTGP and its components}
	\label{class_components_MTGP}
\end{figure}

\paragraph*{}We have widely presented MTGP statuses in this section. Since this generator utilizes a PRNG per block of threads approach, we will need a different status per block to ensure the independence between the random streams produced. The sample program furnished by Saito takes a number of blocks to use as an argument and owns a set of 128 different statuses to feed these blocks. In the next section, we present the protocol that helped us to issue a large number of statuses for MTGP users.

\section{MTGP BENCHMARK}
	\subsection{Empiric Test of 10000 Statuses}
\paragraph*{}We formerly introduced the DC tool, enabling us to create independent PRNGs to use in a parallel environment. It has supplied us 10000 independent parameter sets (parameterized statuses), each corresponding to a different MTGP. This step can be very computationally expensive when the algorithm has to look for generators displaying huge periods, such as $2^{19937}$ for the original MT. According to \citep{LEcuyer2010}, periods contained between $2^{100}$ and $2^{200}$ should be sufficient for nowadays stochastic applications. Thus, we decided to manipulate generators of the lowest period allowed by the MTGP DC, which is still $2^{3217}$. Moreover, the lower the period is, the fewer it wastes bits to store the internal state vector of the PRNG, helping us to save some GPU memory. With this configuration, we have been able to get our entire ready to use MTGPs in a single day, using a 256-node Opteron-enabled Linux cluster.

\paragraph*{}The second phase consisted in applying the BigCrush test battery to each newly created generator, in order to check their quality. First of all, to easily analyze such an amount of results, we modified the TestU01 library output to enable it to produce lighter results output files. In this manner, we have been able to parse results files using script tools like \textit{Sed} or \textit{Awk} to generate statistics. Moreover, since lots of our computations have taken place on the European computing Grid Infrastructure (EGI), we reduced the quantity of data transferred on this slow bandwidth storage space. The use of this HPC tool was compulsory in our case, in fact \citep{LEcuyer.Simard2009} forecasts BigCrush to take about 8 hours of CPU time on an average 64-bit processor. We could not afford to perform the equivalent of 80000 CPU-hours on a single cluster to get our results in a decent time.
\paragraph*{}Our final aim is to provide verified parameterized statuses to allow the simulation community to initialize GPU-enabled PRNGs without introducing any bias in stochastic simulations, according to the current knowledge. A basic selection consisted in keeping only statuses that had perfectly passed all tests of the battery. But given that this approach eliminated approximately 40\% of the statuses, we tried to determine whether other statuses could be kept with a good confidence level. We set up a more formal analysis to answer this question.

	\subsection{Statistics-Based Analysis}
\paragraph*{}Each test of the TestU01 Bigcrush battery \citep{LEcuyer.Simard2007} is governed by the H$_{0}$  hypothesis, that the successive output values of the RNG are i.i.d. U(0, 1), i.e. are independent random variables from the uniform distribution over the interval [0;1]. These tests are defined by a test statistic $\gamma$ (which is a function of the numbers to be tested). They compute and report a number, called the p-value of the test, which is contained between 0 and 1. Furthermore, if $\gamma$ has a continuous distribution, the p-value is U(0, 1) under H$_{0}$. At this point, let us consider two precisions from L’Ecuyer and Simard:
\begin{enumerate}
\item{« If the p-value is extremely small (e.g., less than 10$^{-10}$, then it is clear that the RNG fails the test, whereas if it is not very close to 0 or 1, no problem is detected by this test. » \citep{LEcuyer.Simard2007};}
\item{« Moreover, when a generator starts failing a test decisively, the p-value of the test usually converges to 0 or 1 exponentially fast as a function of the sample size when the sample size is increased further. » \citep{LEcuyer.Simard2009}.}
\end{enumerate}

\paragraph*{}According to these quotations, we decided to consider three p-value types, detailed hereafter:
\begin{itemize}
\item{p-values contained between [0.001;0.999] are reckoned as correct, (these values are proposed by the TestU01 library);}
\item{those included in the range [0;0,001[ U ]0.999;1] are counting as suspect;}
\item{lastly, we refined the previous range since we needed to take into account extremely small p-values (less than 10$^{-10}$), called disastrous afterwards.}
\end{itemize}

\paragraph*{}As mentioned previously, we ran our tests on 10000 independent statuses, with regards to MTGP DC. The column chart appearing on figure \ref{result_table_10000_tests} represents the number of suspect p-values noticed for statuses where no disastrous p-values were obtained (due to layout considerations, only the thirty-two first tests are present on figure \ref{result_table_10000_tests}):

\begin{figure}[H]
        \centering
	\addtolength{\leftskip}{-0.2cm}
	\includegraphics[keepaspectratio, scale=0.28]{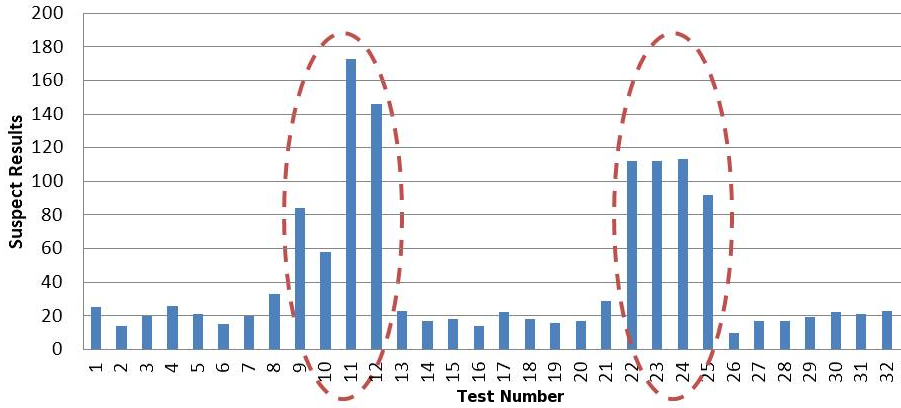}
	\caption{Number of suspect results versus test numbers (extract displaying tests 1 through 32)}
	\label{result_table_10000_tests}
\end{figure}

\paragraph*{}The latter figure helped us to easily identify three test groups. They distinguish from others by recording more than 40 suspect p-values. The interesting point here is that these three groups characterize some aspects of the generator behavior. In fact, tests belonging to the same group are just differently parameterized versions of the same test. The noticeable tests are described as follows in the TestU01 documentation \citep{LEcuyer.Simard2009}:

\begin{itemize}
\item{\textit{smarsa\_CollisionOver} (tests 9 to 12), is an overlapping pairs sparse occupancy (OPSO) test introduced in \citep{Marsaglia1985};}
\item{\textit{snpair\_ClosePairs} (tests 22 to 25), is a m-nearest-pairs (m-NP) test \citep{LEcuyer.Simard2009};}
\item{\textit{swalk\_RandomWalk1} (tests 74 to 79), applies simultaneously several tests based on a random walk of length \textit{l} over the integers, for several (even) values of \textit{l} \citep{LEcuyer.Simard2009}.}
\end{itemize}

\paragraph*{}Under the H$_{0}$' hypothesis of a uniform distribution of the p-values over the interval [0;1], the distribution of the number of suspect p-values is binomial with parameters n = 10000 and p = 2/1000. So, we can reject the H$_{0}$' hypothesis with a very good confidence level (about 7.10$^{-6}$ for each test that accumulates more than 40 suspect p-values. However, we cannot do the same with H$_{0}$. To do so, the $\gamma$-statistic used by the test should present a continuous distribution, which is not the case for the considered tests. Concretely, it means that pointing out suspect p-values brings useful information, but no matter the excesses of suspect p-values, such results do not provide a formal statistic proof of the test failure. That is why we use to consider a test is failed only when it returns disastrous p-values.

\paragraph*{}Six tests are missing on the previous figure: four, the 70, 71, 80, 81, were introduced in our MT description as problematic tests for any MT-like generator. So we increased the execution speed of the test battery simply by disabling those tests that would have systematically produced disastrous p-values. The other two tests, numbered 35 and 100, are the problematic ones. Only these assessments issue disastrous p-values in a non negligible quantity. We noted that about a tenth statuses were failing the $100^{th}$ Test, while more than a fifth went wrong with the $35^{th}$ Test. Figure \ref{results_35_100_mtgp3217} gives a graphical representation of the announced proportions:

\begin{figure}[H]
	\centering
	\includegraphics[width=0.5\textwidth]{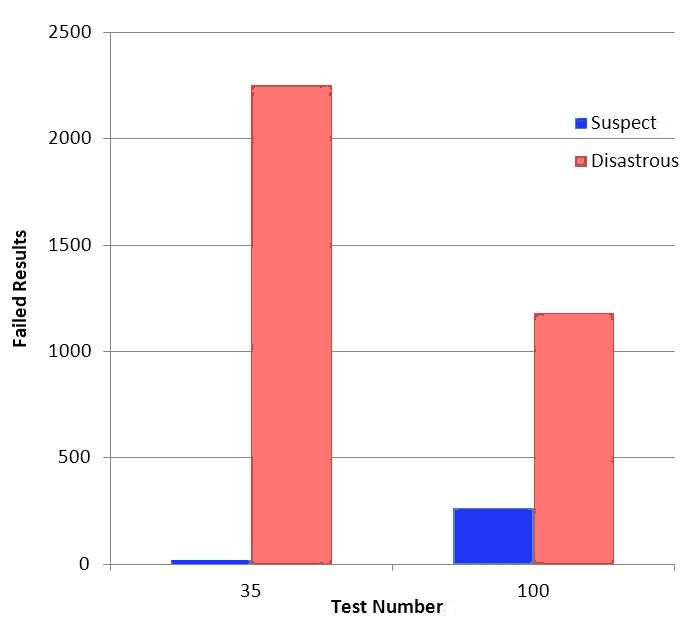}
	\caption{Detailed Results for Tests 35 and 100 of the BigCrush battery}
	\label{results_35_100_mtgp3217}
\end{figure}

\paragraph*{}Test number 35 is the \textit{sknuth\_Gap} test with N = 1, n = $3.10^{8}$, r = 25, Alpha = 0 and Beta = 1/32 \citep{Knuth1969}. This test counts, for s = 0, 1, 2, … « the number of times that a sequence of exactly s successive values fall outside the interval [Alpha, Beta] (this is the number of gaps of length s between visits to [Alpha, Beta]). It then applies a chi-square test to compare the expected and observed number of observations. » \citep{LEcuyer.Simard2009}. A typical generator miscarrying this test « wanders in and out of [Alpha, Beta] for some time, then goes away from [Alpha, Beta] for a long while, and so on » \citep{LEcuyer.Simard2007}. However, one should note that the same test is perfectly passed with other Beta values. In view of the analysis we propose, this fact is obviously logical for Beta values lower than the incriminated one (1/32), but it is rather strange not to find disastrous p-values with a higher Beta value (Test 34 sets Beta to 1/16).
\paragraph*{}The test number 100 is referenced as \textit{sstring\_HammingIndep} test with N = 1, n = 107, r = 25, s = 5, L = 1200 and d = 0. It applies two tests of independence between the Hamming weights of successive blocks of L bits \citep{LEcuyer.Simard1999}. According to François Panneton, this test measures Hamming-weight dependencies between random values issued by a given generator. It tends to demonstrate that the recurrence does not shuffle bits enough from an iteration to another \citep{Panneton2004}.
\paragraph*{}With this first experiment, we have put under the spotlight difficulties encountered by MTGP 2$^{3217}$. Assuming that these problems are mostly concentrated on the two properties checked by tests 35 and 100, those producing disastrous p-values, we have focused our further studies on them.

\section{PARAMETERIZED STATUS INFLUENCE}
	\subsection{Seed Status Variation}
\begin{figure}[H]
	\centering
	\includegraphics[keepaspectratio, scale=0.5]{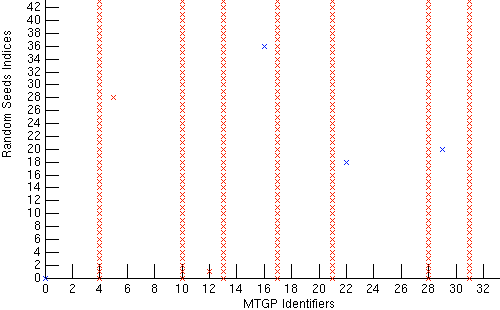}
	\caption{Extracts of the results for Test 35: MTGP identifiers versus random seeds indices}
	\label{screen_test35}
\end{figure}

\paragraph*{}Let us recall that the 10 000 tested statuses were only differing from their parameterized parts, whereas they shared the same seed status, fixed to 0. Viewing the previous results, we decided to work out whether the fact of passing a test or not, and by the way the quality of a generator, was due to either its parameterized status, or its seed status, or both. To do so, we settled a new experiment, sieving a set of 100 parameterized statuses alternately associated with 100 seed statuses randomly chosen. This initialization technique, called Random Spacing, represented a total of 10 000 combinations put to the proof of the guilty tests presented above. Figure \ref{screen_test35} shows an extract of the graphical output for test 35. Red crosses mark a disastrous p-value, while blue ones indicate suspect p-values. An aggregation of red crosses on vertical lines shows that the parameterized status on the abscissa failed the considered test for all the seed statuses it was associated to. So, in the case of parameterized generators, it seems that the parameterized status establishes the generator quality on its own, independently from the selected seed status.

\vspace{-0.2in}

	\subsection{Period Variation}
\paragraph*{}To comfort our previous assertion, we tried to make other elements of the parameterized status vary to observe their impact on the quality of the generator. As long as DC tries to figure out a tempering matrix such as the PRNG produces a well distributed sequence \citep{Matsumoto.Nishimura2000}, modifying this period should highly impact this property for the newly created statuses. Now, we previously brought forward that Test 35 (\textit{sknuth\_gap}) of BigCrush based its judgment on this characteristic. So, we intended to obtain much better results using 1000 MTGPs of period 2$^{23209}$, confronted to tests 35 and 100 only. The result is crystal-clear since 99.5\% of the statuses passed both tests without any problem. Moreover, we only noticed suspect p-values in the other 0.05\%.
\paragraph*{}Obviously, a higher period eliminates sequence distribution issues, but this latter result could hide potential intrinsic weaknesses of the MTGP algorithm. Our last experiment will introduce as a standard a quality-proven PRNG of the same family: the original MT \citep{Matsumoto.Nishimura1998}.

	\subsection{Algorithm Variation}
\paragraph*{}The original MT is designed with linear-recurrences preventing its recommendation for some particular applications such as cryptography. As far as we know, no other problems are referenced concerning this PRNG. That is why we consider it as a good standard to compare with the target of our investigations. An interesting point is that its own DC is able to produce parameters for the 2$^{3217}$ period, thus allowing us to work with MTs and MTGPs of the same period. This way, we focus our experiment on the algorithm-dependent parts of the parameterized statuses. Once again, we observed the behavior of each generator when faced to tests 35 and 100. We selected a sample of 1000 independent PRNGs to compare outputs with our previous benchmarks. Results are even more clear-cut than before: 99.9\% of MT-dedicated statuses passed the two tests without any failure, whereas about 64\% of MTGP statuses did.
\paragraph*{}The two previous results tend to show the influence of parameterized statuses. Here we have shown that this data structure is tightly bound to the algorithm using it. MTGP seems to present weaknesses when configured with shorter periods, since MT, running with the same relatively small period, eschews traps where its recent variant falls frequently into. This section results are summed up on the column chart displayed on figure \ref{results_35_100_all}:
 
\begin{figure}[H]
	\centering
	\includegraphics[keepaspectratio, scale=0.76]{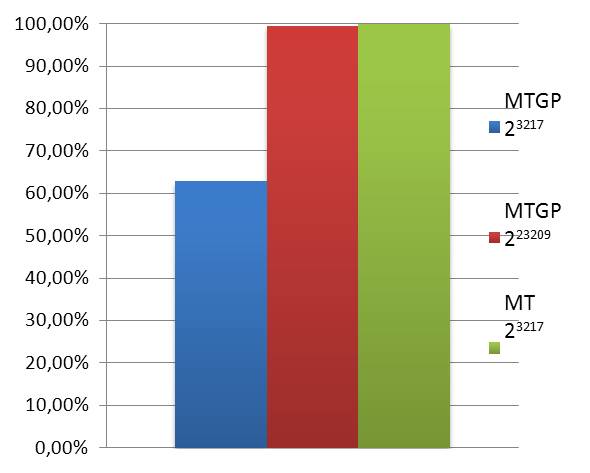}
	\caption{Percentage of passed results noticed for tests 35 and 100 depending on the PRNG}
	\label{results_35_100_all}
\end{figure}
\vspace{-0.2in}

\section{CONCLUSION}
\paragraph*{}This work was first intended to study a particular generator designed for GPU hardware architectures: MTGP. After an introduction to other recent approaches mentioned in the literature, in order to compare them with the MTGP strategy, we provided a description of MTGP and proposed a generic object model representing generators using distinct seeds and parameters. To complete our description, we achieved some analysis of this PRNG by facing it to the current most stringent test battery: BigCrush from the TestU01 test software library. Weaknesses identified during these experiments have been reduced by comparisons with other configurations of the generator as well as with the original MT.

\paragraph*{}The second goal of this study was to furnish parameterized statuses to the simulation community, to allow them to quickly and safely initialize their MTGPs without introducing any potential bias with a bad status. Only statuses that have passed BigCrush have been kept and will be freely available on the Internet.

\paragraph*{}We have shown that MTGP was safer with longer periods, according to TestU01 criteria, but the more a PRNG period is long, the more space it needs to store the internal state vector used by its algorithm. Nowadays, GPUs memory characteristics do not allow us to waste bytes to store PRNG data without influencing the whole application speed. By selecting only parameterized statuses referenced by our study or proofed by an equivalent benchmarking protocol, scientists using MTGP with the lowest available period (i.e. 2$^{3217}$) can dramatically reduce the memory footprint of their hybrid stochastic simulations.

\paragraph*{}However, we should not forget that the empirical tests provided by TestU01 consider random sequences individually. So, we are now thinking of a way to test mixed random sequences, the way they can be in a single GPU application using several blocks of threads fed with different sources of randomness.

\bibliographystyle{apalike}
\bibliography{esm2010}

\section{BIOGRAPHY}
\paragraph*{\textsc{JONATHAN PASSERAT-PALMBACH}} is a PhD student at the LIMOS (ISIMA) - UMR CNRS 6158 of Blaise Pascal University of Clermont Ferrand (France). His work, is focused on high performance computing tools and discrete events simulation, applied to ecological modeling.\\
email: \texttt{passerat@isima.fr}

\paragraph*{\textsc{CLAUDE MAZEL}} is an associate professor at the ISIMA Computer Science and Modelling Institute, where he currently manages the Software Engineering and Computing Systems Department. His main scientific interests concern modelling, discrete event simulation, and design methods for simulation software, applied to ecological modelling.\\
email: \texttt{mazel@isima.fr}

\paragraph*{\textsc{ANTOINE MAHUL}} is currently an engineer in the field of scientific computing and managing of HPC and grid resources for universities of Clermont-Ferrand. He obtained his PhD in computer science at the LIMOS of Blaise Pascal University in 2005.\\
email: \texttt{antoine.mahul@univ-bpclermont.fr}

\paragraph*{\textsc{DAVID HILL}} is currently Vice President of Blaise Pascal University in charge of Computer Science and past co-director of ISIMA Computer Science \& Modeling Institute. Since 1990, Professor Hill has authored or co-authored more than a hundred technical papers and journal papers and he has published various text books including free e-books from Blaise Pascal University Press (\url{http://www.isima.fr/~hill}).\\
email: \texttt{david.hill@univ-bpclermont.fr}

\end{document}